\documentclass[aps,prd,twocolumn,reprint,preprintnumbers,amsmath,amssymb,nofootinbib,superscriptaddress]{revtex4}
\usepackage{graphicx}
\usepackage{subfigure}
\usepackage{epsfig}
\usepackage{dcolumn}
\usepackage{bm}
\usepackage{ulem}
\usepackage{color}
\usepackage{multirow}
\usepackage{slashed}
\usepackage{hyperref}
\usepackage{amsthm}

\def\avg#1{\left\langle#1\right\rangle}
\def\bra#1{\left\langle#1\right|}
\def\ket#1{\left|#1\right\rangle}

\def\kc#1{\left(#1\right)}
\def\kd#1{\left[#1\right]}
\def\ke#1{\left\{#1\right\}}

\def\be{\begin{equation}}       \def\ee{\end{equation}}
\def\bea{\begin{eqnarray}}      \def\eea{\end{eqnarray}}
\def\ba{\begin{array}}
    \def\ea{\end{array}}
\def\bnum{\begin{enumerate} }
    \def\enum{\end{enumerate}}
\def\bpm{\begin{pmatrix}[0.5]}
    \def\epm{\end{pmatrix}}
\def\bbm{\begin{bmatrix}[0.5]}
    \def\ebm{\end{bmatrix}}

\def\nn{\nonumber}

\def\=>{\Rightarrow}
\def\>{\rightarrow}

\def\eye2{Fathbb{I}}

\def\Tr{\mathrm{Tr}}

\renewcommand{\>}{\rangle}

\makeatletter
\renewcommand*\env@matrix[1][\arraystretch]{%
    \edef\arraystretch{#1}%
    \hskip -\arraycolsep
    \let\@ifnextchar\new@ifnextchar
    \array{*\c@MaxMatrixCols c}}
\makeatother

\begin{document}

    \newcommand{\IHEP}{\affiliation{Institute of High Energy Physics, Chinese Academy of Sciences, Beijing 100049, China}}
    \newcommand{\UCAS}{\affiliation{School of Physics, University of Chinese Academy of Sciences, Beijing 100049, China}}
    \newcommand{\ITP}{\affiliation{Institute of Theoretical Physics, Chinese Academy of Science, Beijing 100190, China}}

    \title{Quantum error correction and entanglement spectrum in tensor networks}
    \author{Yi Ling}
    \email{lingy@ihep.ac.cn}
    \author{Yuxuan Liu}
    \email{liuyuxuan@ihep.ac.cn}
    \IHEP\UCAS
    \author{Zhuo-Yu Xian}
    \email{xianzy@itp.ac.cn}
    \IHEP\UCAS\ITP
    \author{Yikang Xiao}
    \email{ykxiao@ihep.ac.cn}
    \IHEP\UCAS

    \begin{abstract}
A sort of planar tensor networks with tensor constraints is
investigated as a model for holography. We study the greedy
algorithm generated by tensor constraints and propose the notion
of critical protection (CP) against the action of greedy
algorithm. For given tensor constraints, a CP tensor chain can be
defined. We further find that the ability of quantum error
correction (QEC), the non-flatness of entanglement spectrum (ES)
and the correlation function can be quantitatively evaluated by
the geometric structure of CP tensor chain. Four classes of tensor
networks with different properties of entanglement is discussed.
Thanks to tensor constraints and CP, the correlation function
is reduced into a bracket of Matrix Production State and the
result agrees with the one in conformal field theory. 

    \end{abstract}
    \maketitle

\section{Introduction}
    \label{SectionInt}
Quantum entanglement plays a key role in understanding the
structure of spacetime from the emergent point of view
\cite{Maldacena:2001kr,VanRaamsdonk:2010pw}. The Ryu-Takayanagi
(RT) formula links the entanglement entropy of a subsystem on the
boundary to the area of the minimal homological surface in the
bulk \cite{Ryu:2006bv}. Such an approach has been recently
generalized to construct the gravitational dual of Renyi entropy
\cite{Dong:2016fnf}, which provides a correspondence of
entanglement spectrum (ES) between the bulk and the boundary. In
particular, for the vacuum in $AdS_3/CFT_2$ correspondence, Renyi
entropy satisfies Cardy-Calabrese formula and a non-flat ES is
inherent \cite{Calabrese:2004eu,Calabrese:2009qy}. Another
remarkable feature of AdS space is the subsystem duality, which
states that a local operator in the bulk can be reconstructed in a
subsystem $A$ on the boundary if it is located within the
entanglement wedge of $A$
\cite{Hamilton:2006az,Dong:2016eik,Cotler:2017erl,Freivogel:2016zsb,Mintun:2015qda,Almheiri:2014lwa,Harlow:2018fse}.
It can be be viewed as the accomplishment of Quantum Error
Correction (QEC) in quantum information
\cite{Schumacher:1996dy,Almheiri:2014lwa,Pastawski:2015qua,Harlow:2018fse}.
Moreover, it is found that RT formula can be derived from QEC
\cite{Harlow:2016vwg}.

It has been revealed that tensor networks provide a geometric
picture for entanglement renormalization such that holographic
spaces may emerge from the entanglement of a many-body
system\cite{Vidal:2007hda,Swingle:2009bg,Swingle:2012wq}, gearing
up the exploration on the deep relation between tensor networks
and the structure of spacetime \cite{Nozaki:2012zj,Qi:2013caa}.
One typical kind of tensor networks is the multiscale entanglement
renormalization ansatz (MERA), which respects RT formula,
exhibiting logarithmic law of entanglement entropy and non-flat
entanglement spectrum as the AdS vacuum
\cite{Vidal:2007hda,Swingle:2009bg,Swingle:2012wq,Kim:2016wby,Bao:2015uaa}.
However, MERA breaks the isometry group $SL(2,R)$ and has a
preferred direction, implying that QEC can not be realized along
all directions. On the other hand, perfect tensors, which are also
called as holographic codes, take the advantages of implementing
QEC over a $H^2$
space \cite{Pastawski:2015qua,Yang:2015uoa,Donnelly:2016qqt}.
Unfortunately, it is found that such kind of tensor networks has a
flat ES and trivial connected correlation functions, which
evidently is not a reflection of the holographic property of AdS
spacetime \cite{Bhattacharyya:2016hbx,Evenbly:2017htn}. In random
tensor networks and spin networks, all the orders of Renyi entropy
for the ground state share the same RT formula, leading to a flat
ES as
well \cite{Hayden:2016cfa,Qi:2018shh,Chirco:2017wgl,Han:2016xmb}.
The attempt to recover the result of Cardy-Calabrese formula of
Renyi entropy, $S_n=(1+1/n)(c/6) \log(l/\epsilon)$, can be found in \cite{Han:2017uco} where the bulk
dynamics is taken into account.

Recently a new class of tensor networks which is named
hyperinvariant tensor networks has been constructed in
\cite{Evenbly:2017htn}, which retains the advantages of both MERA
and perfect tensor networks. The key ingredient of hyperinvariant
tensor networks is to impose multi-tensor constraints, which
demand certain product of multiple tensors to form an isometric
mapping. Remarkably, this sort of tensor networks can not only
accomplish QEC as perfect tensors, but also generate non-flat ES
as MERA, thus qualitatively capturing both holographic features of
AdS spacetime.

Nevertheless, some key issues remain unanswered in this approach.
First of all, the holographic property of tensor networks depends
on the specific structure of tensor constraints. What kind of
multi-tensor constraints could endow desirable features of AdS
spacetime to a given tensor network? More importantly, to
accomplish the holographic features of tensor networks one always
faces a dilemma: once the ability of QEC of a tensor network
becomes stronger, then more easily its ES becomes flat, and vice
versa. Is there any criteria to characterize the ability of QEC
and the non-flatness of ES for a tensor network with given
constraints? At the same time, can any feature of CFT be
reflected by the specific structure of tensor networks? We wish
to answer above issues based on some examples of tensor network.

We will construct tensor networks by tiling $H^2$ space with
identical polygons, and then impose tensor constraints with the
notion of tensor chain, which leads to a generalized description
of greedy algorithm. We will investigate QEC, ES and correlation
function by manipulating tensor networks. For the ES and correlation function, we will also compare our holographic results with the results in conformal field theory. Moreover, we will propose the notion of critical protection (CP) to describe the
behavior of tensor networks under the greedy algorithm. A
geometric quantity $\kappa_c$, named as the average reduced
interior angle of a tensor chain, will be proposed to measure the ability of QEC and justify the flatness
of ES.

\section{Constraints on tensor chains}
\subsection{Tensor Chains}
We discretize $H^2$
space uniformly by gluing identical polygons composed of $b$
edges, with $a$ edges sharing the same node. We call such
discretization as the $\{b,a\}$ tiling of $H^2$ space. Since the
sum of interior angles of a triangle in a space with negative
curvature must be less than $2\pi$, a $\{b,a\}$ tiling of $H^2$
space can be realized only if
    $\frac1a+\frac1b<\frac12$.

\begin{figure}
        \newcommand{\minipagewidth}{0.47\linewidth}
        \begin{minipage}[t]{\minipagewidth}
            \centering
            \includegraphics[width=0.9\linewidth]{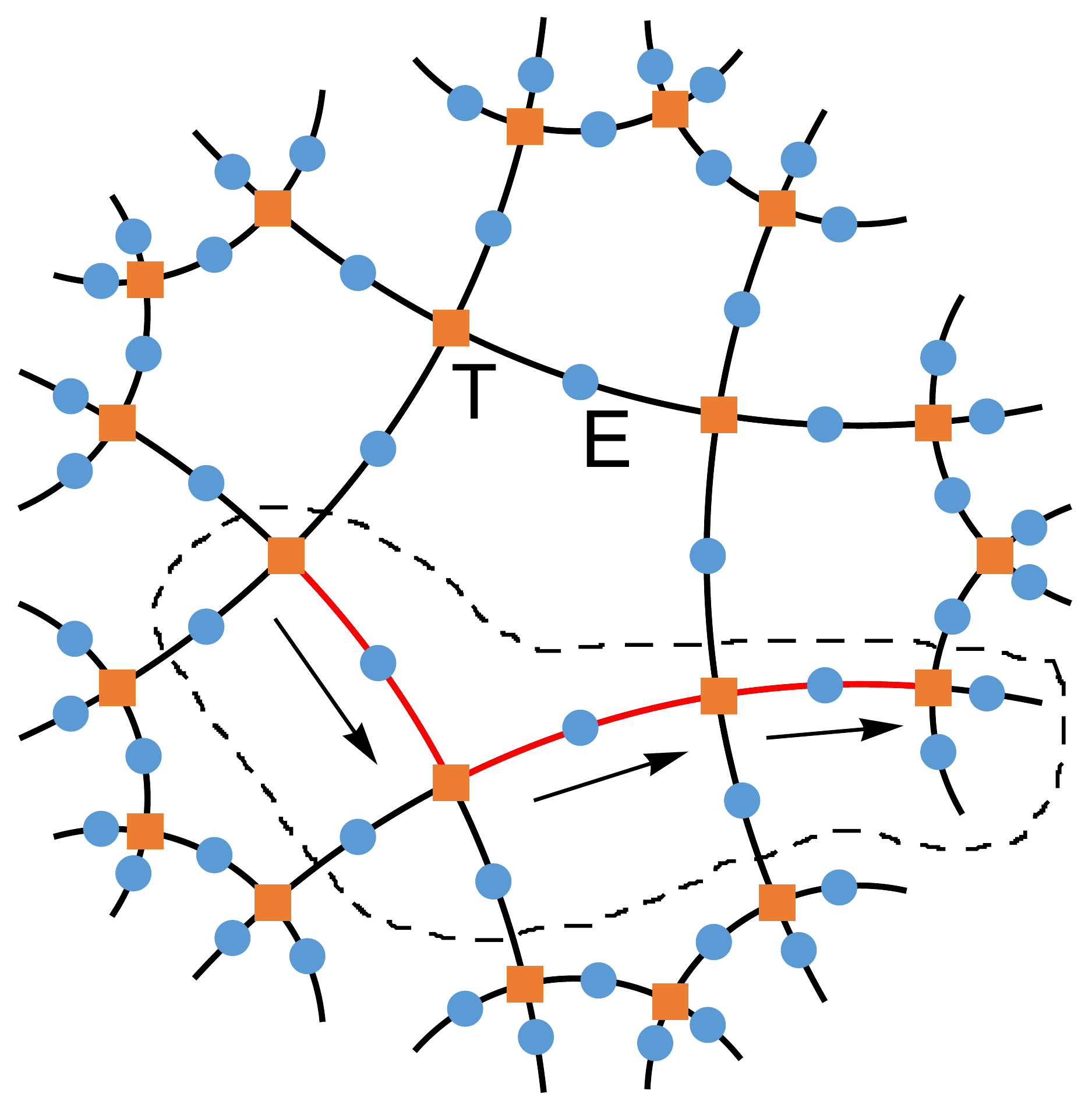}
            \caption{A tensor network with $\ke{5,4}$ tiling. Enclosed by the dashed line is an example of tensor chain. Its skeleton forms a directed polyline which is marked in red.}\label{Figtiling}
        \end{minipage}
        \hfill
        \begin{minipage}[t]{\minipagewidth}
            \centering
        \includegraphics[width=1\linewidth]{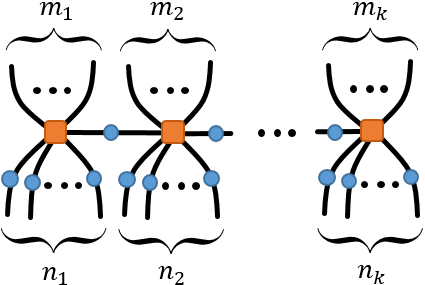}\\
        \caption{A general form of tensor chain.}\label{FigTC}
        \end{minipage}
    \end{figure}

A tensor network can be constructed based on each $\{b,a\}$
    tiling, as illustrated in Fig.~\ref{Figtiling}. Associated with each node,
    we define a tensor $T$ with $a$ indexes, each of which is specified to an
    edge jointed at the node respectively. Associated with
    each edge, we define a tensor $E$ with $2$ indexes.
    Because of the rotational invariance of $H^2$ space, we demand that the indexes of tensor $T$ and $E$ have cyclic
    symmetry
    \be
    T^{i_1i_2\cdots i_a}=T^{i_2i_3\cdots i_ai_1},\quad  E_{i_1i_2}=E_{i_2i_1}.
    \ee
Consider a tensor network $\Psi$, and let all the indexes of
tensors $T$ contract with those of tensors $E$ such that all
uncontracted indexes belong to tensors $E$ only. Corresponding to
such a network, we define a state $\ket{\Psi}$ in the Hilbert
space on those uncontracted edges.

By dissecting a tensor network, as in Fig.~\ref{Figtiling}, we
define a key object called tensor chain $M$, whose general form is
shown in Fig.~\ref{FigTC}. Vividly, the uncontracted edges in $M$
are split into the upper part $A$ and lower part $B$. So we
denote its elements as $M^A_B$. The number of edges at each node
satisfies $m_i+n_i=a-2+\delta_{i1}+\delta_{ik}$, where $i$ is
the sequence number labelling the node of tensor chain.
Specifically, for the tensor chain in Fig.~\ref{Figtiling},
$k=4,\,\kc{m_1,m_2,m_3,m_4}=\kc{2,0,1,1},\,\kc{n_1,n_2,n_3,n_4}=\kc{1,2,1,2}$.

Vice versa, a tensor chain can be mapped into the tiling of $H^2$
space and its skeleton forms a directed polyline in the
network, where along the direction of the polyline the sequence
number $i$ increases and the upper (lower) edges are placed on the
left (right) hand side of the polyline. To describe the curvature
of its corresponding polyline, we define the average reduced
interior angle of a tensor chain as \be
 \kappa=\frac1k\kc{\sum_{i=1}^k m_i + k-1},
\ee
where ``reduced'' means that we have taken ${2\pi\over a}$
as the unit of interior angles.

We will focus on the tensor network
with $\ke{5,4}$ tiling as a typical example to disclose the
structure of the tensor chain which is critically protected under the action of greedy algorithm.
Our analysis and results can be generalized to the tensor networks with $\ke{a,b}$ tiling.

\subsection{Tensor Constraints}\label{SectionTC}

\subsubsection{Tensor network with $\kappa_c=2$}
We define tensor constraints as follows. Besides the cyclic
symmetry, we further impose constraints on rank-$4$ tensor $T$
(orange square) and rank-$2$ tensor $E$ (blue circle), such that
they satisfy the following equations
\begin{equation}
    \includegraphics[height=0.06\textheight]{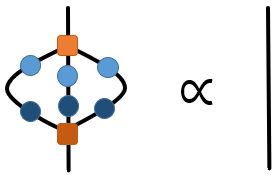}
    \quad , \quad
    \includegraphics[height=0.06\textheight]{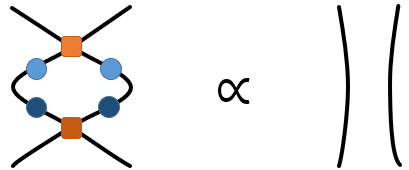},
\end{equation}
where the conjugation of tensors are marked in dark colors. In
other words, the tensor chains
\begin{equation}\label{542PITC}
    \includegraphics[height=0.05\textheight]{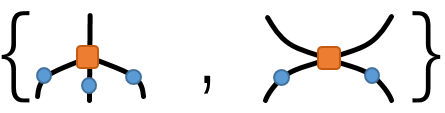}.
\end{equation}
are proportional to isometries from the Hilbert space on upper
edges to the Hilbert space on lower edges. For convenience, in
the remainder of this paper we will adopt the expression like
(\ref{542PITC}) to represent tensor constraints for short. The
shape of the tensor chain in constraints (\ref{542PITC}) can be
characterized by their average reduced interior angle, which is
$\ke{1/1=1,~2/1=2}$. We will call the maximal one as the CP
reduced interior angle $\kappa_c=2$, as the reason will become
clear later.

For the simple case as illustrated in (\ref{542PITC}), each of tensor chains only involves a single tensor $T$. One can derive other tensor chains proportional to isometries as well from the tensor constraints.
For instance, from (\ref{542PITC}), the
following tensor chains are proportional to isometries.
\be
\includegraphics[height=0.05\textheight]{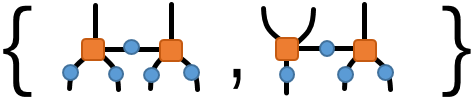}.
\ee

The detailed analysis is given in \cite{Ling:2018ajv}. Here we
just argue that all of these tensor chains form a set $S_D$ with
infinite number of elements, and satisfy $\kappa\leq\kappa_c$. The subscript $D$ refers to the fact that $S_D$ is derived from tensor constraints given. We
stress that one should take all these tensor chains into account
when justifying whether the contraction of tensor product
could be simplified under the action of greedy algorithm.

We further require that any tensor chain which is proportional to
isometry can be derived from tensor constraints, which
restricts the structure of tensor $T$ and $E$. In other words, we
require that those tensor chains which do not belong to the set
$S_D$ should {\it not} be propositional to isometries, which
prevents tensor $T$ and $E$ from trivial structure, for instance,
the outer product of identity matrices. We point out that many
tensor chains do not belong to $S_D$, such as the following tensor
chains for constraints (\ref{542PITC}) \be\label{542NPITC}
\includegraphics[height=0.05\textheight]{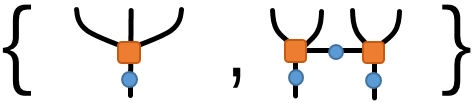},
\ee

\subsubsection{Tensor network with $\kappa_c=3/2$}
Definitely, we may impose other tensor constraints, for
instance, by requiring
the following tensor chains to be proportional to isometries.
\begin{equation}
\includegraphics[height=0.05\textheight]{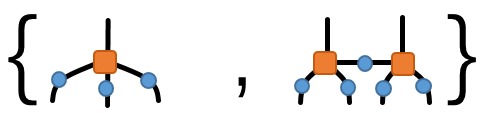},
\label{5411PITC}
\end{equation}
whose average reduced interior angles are $\ke{1,3/2}$, then
$\kappa_c=3/2$. Similarly, from (\ref{5411PITC}), the following
tensor chains are proportional to isometries \be
\includegraphics[height=0.05\textheight]{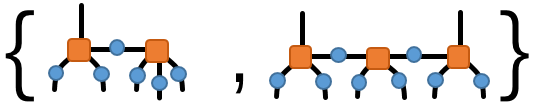},
\ee 
but the following tensor chains are not \be
\includegraphics[height=0.05\textheight]{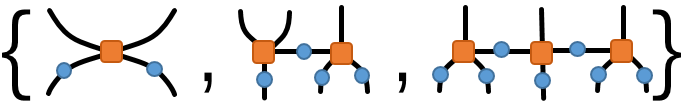}.
\ee

The specific construction of tensors $T$ and $E$ subject to
above constraints is given in Appendix \ref{SectionExistence}.

\section{Greedy Algorithm and Protection}

\subsection{General Greedy Algorithm with Tensor Chain}\label{SubSectionPTNHTN}
We firstly review the greedy algorithm on a tensor network
    following the description in \cite{Pastawski:2015qua}, which
    provides an intuitive way to figure out the region in which the
    corresponding sub tensor network must be an isometry. Beginning
    with an interval $A$ on the boundary of a tensor network $\Psi$,
    we consider a sequence of cuts $\ke{C_n}$, each of which is
    bounded by $\partial A$ and obtained from the previous one by a
    local move on the lattice. The corresponding sub tensor networks
    also form a sequence of $\ke{\Phi_n}$, where $\Phi_n$ consists of
    those tensors between $A$ and $C_n$. Let $C_1=A$ and $\Phi_1$ is an identity. For perfect tensors, at each step one figures out a tensor $M_n$ which has at least half of its legs contracted with $\Phi_n$ and construct $\Phi_{n+1}$ by adding $M_n$ to $\Phi_n$ such that $\Phi_{n+1}$ must be an isometry as well. The procedure stops when one fails to add such tensors to the sequence.

We can generalize the above description by replacing its single
    tensor $M_n$ by a tensor chain $M_n$ which is proportional to
    isometry, with lower edges contracted with $\Phi_n$. According to section \ref{SectionTC}, those tensor chains which are proportional to isometries
    form a set $S_D$ derived from tensor constraints.

Furthermore, we can generalize the target of greedy algorithm to a tensor chain $M$ rather than a tensor network $\Psi$. Given a
    tensor chain $M$, we simplify the contraction $\sum_B
    M_B^A(M_B^C)^*$ subject to tensor constraints. For example,
    according to (3), we can simplify the contraction
    \be\label{542GA31}
    \includegraphics[height=0.08\textheight]{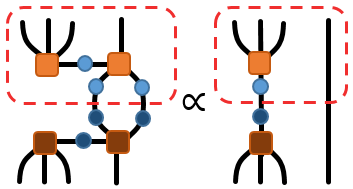}.
    \ee
Similarly, according to (5), we can simplify the contraction
    \be\label{5411GA1003}
    \includegraphics[height=0.08\textheight]{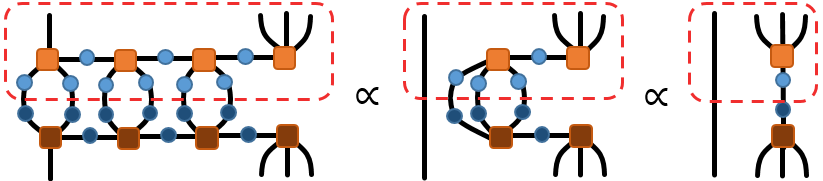}.
    \ee

\begin{figure*}
    \centering
    \includegraphics[height=0.18\textheight]{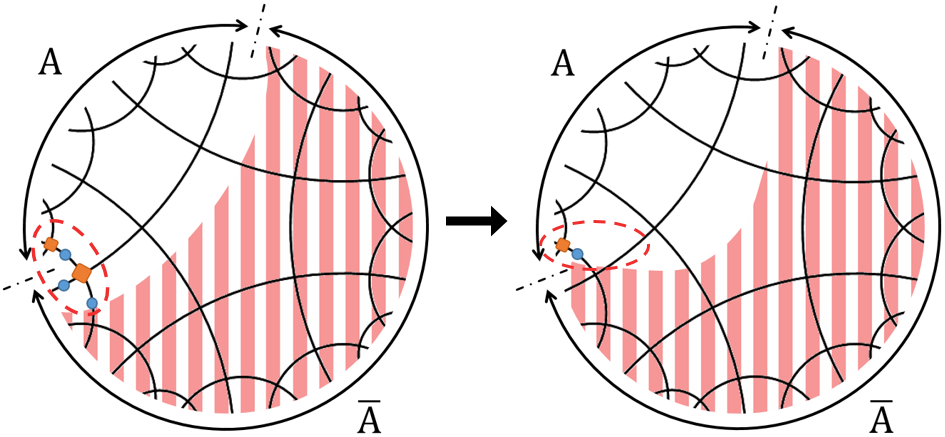}
    \caption{One step in the greedy algorithm generated from (3) beginning at $\bar A$.}
    \label{Fig542GA}
    \includegraphics[height=0.18\textheight]{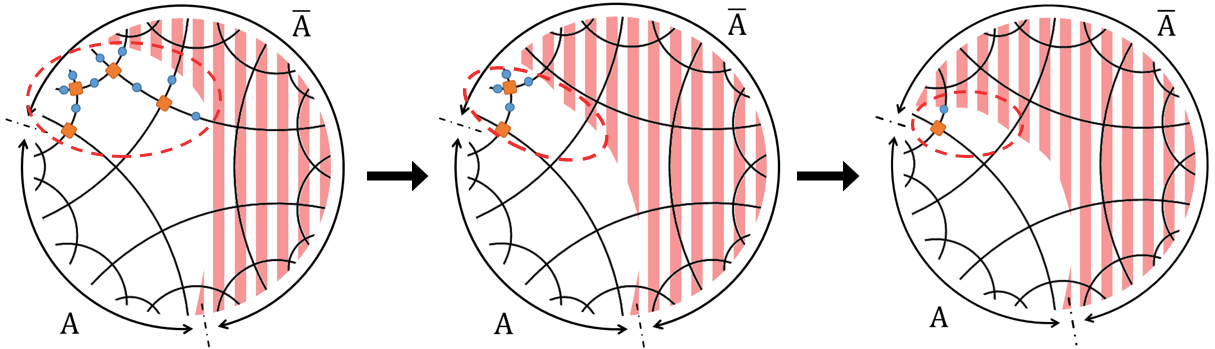}
    \caption{Two steps in the greedy algorithm generated from (5) beginning at $\bar A$.}
    \label{Fig5411GA}
\end{figure*}

Actually, the above description of greedy algorithm is
    equivalent to the description in \cite{Pastawski:2015qua} for a
    single-interval. At each step from $\Phi_n$ to $\Phi_{n+1}$, a
    $M_n\in S_D$ is used to simplify a tensor chain $M$. For example,
    the procedure of simplifying (\ref{542GA31}) corresponds to
    the step of extending the shaded region as illustrated in
    Fig.~\ref{Fig542GA}, where the corresponding tensors are enclosed
    by dashed line in red. Similarly, the process of simplifying
    (\ref{5411GA1003}) corresponds to those steps in
    Fig.~\ref{Fig5411GA}.

We define that a tensor chain $M$ is unprotected if it can be
simplified under the action of greedy algorithm. Otherwise, say it
is protected.

\subsection{Critical Protection (CP)}

Generally speaking, when the tiling and tensor constraints are
given, the larger $\kappa$ is, the easier a tensor chain becomes
unprotected. The protected and endless $M$ with largest $\kappa$
is called critically protected (CP) tensor chain $M_c$.
Equivalently, one can check that $M_c$ would become unprotected
once the list of its $n_i$ are rearranged or increased. Its
$\kappa$ is called CP reduced interior angle $\kappa_c$.

\begin{figure}
    \newcommand{\minipagewidth}{0.48\linewidth}
    \newcommand{\figheight}{120pt}
    \begin{minipage}[t]{\minipagewidth}
        \centering
        \includegraphics[height=\figheight]{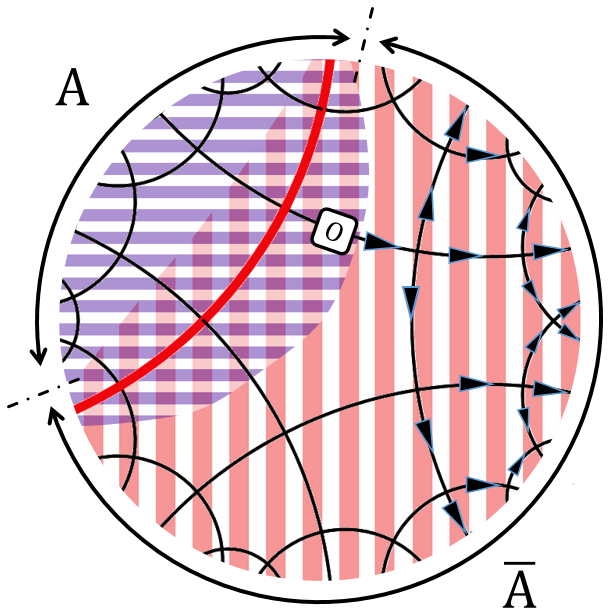}
        \caption{The tensor network with $\{5,4\}$ tiling and tensor constraints (\ref{542PITC}). Those tensors within the shaded region with purple (red) stripes is absorbed by the greedy algorithm  starting from $A$ ($\bar A$).
            The CP tensor chain is marked by a solid line in red.
            An operator $O$ in the bulk is pushed to a sub-interval of $\bar A$.
        }\label{Fig542}
    \end{minipage}
    \hfill
    \begin{minipage}[t]{\minipagewidth}
        \centering
        \includegraphics[height=\figheight]{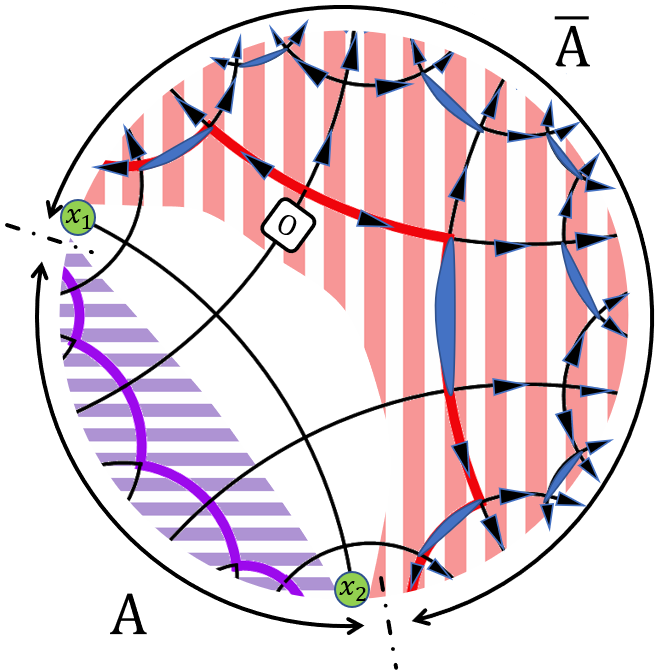}\\
        \caption{The tensor network with $\{5,4\}$ tiling and tensor constraints (\ref{5411PITC}). The tensors in the blank region are not absorbed by the greedy algorithm. Two CP tensor chains correspond to thick polylines in purple and red, respectively.
            An operator $O$ enclosed by the CP tensor chain is pushed to a region within $\bar A$ on the boundary, where blue edges in the shape of rod indicate the employment of the second constraint in (\ref{5411PITC}).
        }\label{Fig5411}
    \end{minipage}
\end{figure}

Under the greedy algorithm generated by (\ref{542PITC}), if
$\exists i$ {\it s.t.} $n_i>1$, then $M$ is unprotected. So CP
tensor chain has the form as plotted on the right hand side of
(\ref{54CP1})
\be\label{54CP1}
\includegraphics[height=0.05\textheight]{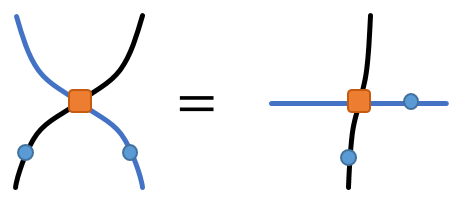}
\includegraphics[height=0.05\textheight]{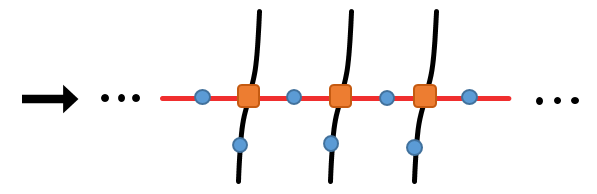}.
\ee Here we have also presented a scheme to figure out CP tensor
chain by a manipulation on the second constraint in
(\ref{542PITC}). The skeleton of such a CP tensor chain forms a
polyline in the tensor network, as shown in Fig.~\ref{Fig542}. For
(\ref{54CP1}), $\kappa_c=2/1=2$, which is just equal to the
maximal one of the average reduced interior angles of the tensor
chains in constraints (\ref{542PITC}).

We deduce the CP tensor chain $M_c$ from constraints (\ref{5411PITC}) as follows. From the first
constraint, we know the number of lower edges
at any node in CP tensor chain should be smaller than three; while
from the second constraint, we know any two nodes with two lower
edges can not be neighbored (otherwise they would be swallowed by
the constraint). Therefore, $M_c$ has the form as plotted on the
right hand side of (\ref{5411CP01})
\be\label{5411CP01}
\includegraphics[height=0.05\textheight]{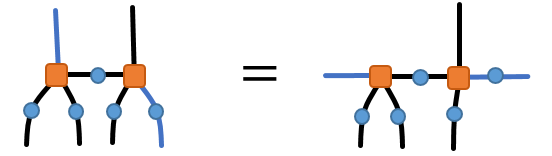}
\includegraphics[height=0.05\textheight]{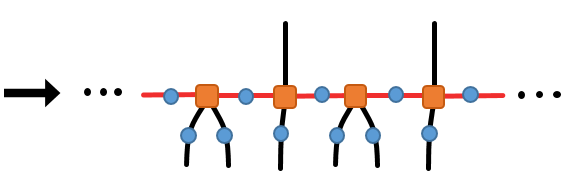}.
\ee
Similarly, one can construct $M_c$ based on the second constraint
in (\ref{5411PITC}), as demonstrated in (\ref{5411CP01}). The
corresponding polyline in the network is marked in
Fig.~\ref{Fig5411}. The CP reduced interior angle is
$\kappa_c=(1+2)/2=3/2$.

\section{QEC and ES}
Throughout this paper we will only consider the QEC by inserting
an operator into the inter bonds, for instance,
    \be\label{insert}
    \includegraphics[height=0.05\textheight]{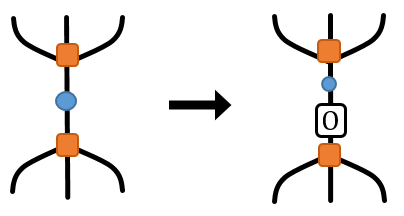}.
    \ee
By virtue of tensor constraints, one can
    push an operator $O$ `through' tensor chains in (\ref{542PITC}) and turn into an operator $O'$, namely

    \newcommand{\figpush}{0.05\textheight}
    \bea
    \includegraphics[height=\figpush]{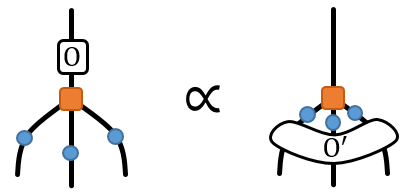}\quad,\quad
    \includegraphics[height=\figpush]{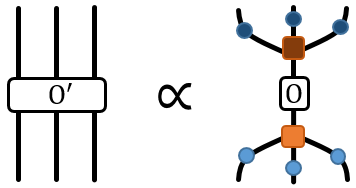},
    \label{1to3}    \\
    \includegraphics[height=\figpush]{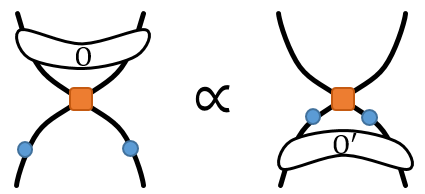}\quad,\quad
    \includegraphics[height=\figpush]{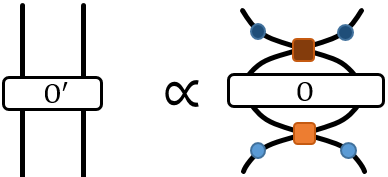},
    \label{2to2}
    \eea
where the conjugation of tensors are marked in dark colors.
Then one can realize the algorithm of QEC, as shown in
Fig.~\ref{Fig542}. Actually, pushing an operator to an interval
$\bar A$ on the boundary is the inverse of the greedy algorithm
beginning at $\bar A$. So any operator inserted outside the CP
tensor chain can be pushed to the boundary.

Next we consider the ES of the reduced density matrix
\begin{eqnarray}
\rho_A=\frac1Z\Tr_{\bar A}{\ket{\Psi}\bra{\Psi}}=\frac1Z \Psi\Psi^\dagger,\quad Z=\avg{\Psi|\Psi},
\end{eqnarray}
where $\bar A$ is contracted in the matrix production of tensor
networks $\Psi$ and $\Psi^\dagger$, and the normalized factor
$Z$ is obtained by contracting all the indexes between them.
$\rho_A$ has a flat ES if all the non-zero eigenvalues are
identical. From the diagonalization of $\rho_A$, we know that the
flatness of ES is equivalent to 
\be\label{flatES}
\rho^2_A\propto\Psi\Psi^\dagger\Psi\Psi^\dagger\propto\Psi\Psi^\dagger\propto\rho_A,
\ee 
which is also equivalent to state that all the orders of Renyi
entropy are equal. If all the tensors are absorbed by the greedy
algorithm starting from $A$ and from $\bar A$ respectively, then
the relation in (\ref{flatES}) holds and leads to a flat ES.
Otherwise the ES is generally non-flat. When there exist
tensors which are not absorbed by the greedy algorithm, although
we can not exclude the tiny possibility that (\ref{flatES})
happens to be valid for some construction of tensor $T$ and $E$
under fine-tuning, we still call that the ES is non-flat for a
general construction of tensor $T$ and $E$.

We show the result of the greedy algorithm acting on the tensor
network with constraints (\ref{542PITC}) in Fig.~\ref{Fig542}.
It indicates that the ES is flat, which coincides with the results
in \cite{Bhattacharyya:2016hbx}. While for the tensor network with
constraints (\ref{5411PITC}), the ES is non-flat as shown in
Fig.~\ref{Fig5411}. At the same time, we point out that the
ability of QEC in this network is weakened in comparison with that
in the network with (\ref{542PITC}), because the operator inserted
into the region enclosed by CP tensor chains will approach the
endpoints of $A$ during the pushing process. Such phenomenon may
be related to the approximate QEC
\cite{Schumacher:2002aqe,Almheiri:2014lwa}.

The boundary effect in above analysis should be stressed. One may
notice that CP tensor chain $M_c$ itself falls into the shaded
region, implying that it is absorbed by the greedy algorithm. This
phenomenon results from the boundary effect in a network with
finite layers, where besides the lower edges of $M_c$, the edges
at the end of $M_c$ need to be contracted as well. The boundary
effect of greedy algorithm is investigated with details in
\cite{Ling:2018ajv}. Here we just remark that this effect is very
limited, only swallowing finite layers (usually only one layer) of
tensors enclosed by $M_c$.

\begin{figure}
    \centering
    \includegraphics[width=0.5\linewidth]{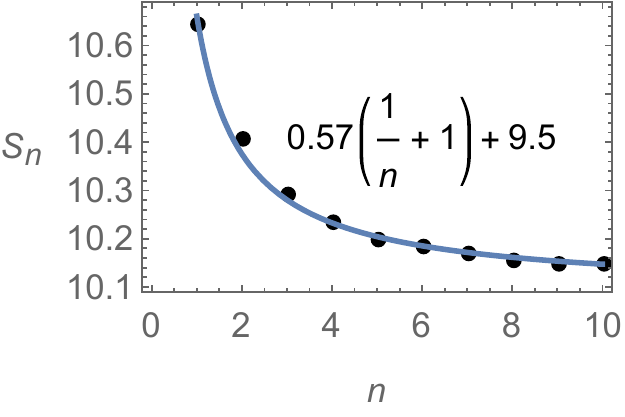}
    \caption{Renyi entropy $S_n$ as a function of $n$, denoted by black dots and fitted by blue line.}\label{FigRenyi}
\end{figure}

Next we investigate the Renyi entropy $S_n$ for the tensor
network composed of the tensors subjected to constraint
Fig.~\ref{5411PITC}. The specific construction of tensors $T$ and
$E$ is given in Fig.~\ref{FigTE5411}, where elementary tensors
$Q$ and $U$ satisfy the relation in (\ref{UQ}). We numerically
calculate $S_n$ for region $A$ in the tensor network in
Fig.~\ref{Fig5411}. The result is shown in Fig.~\ref{FigRenyi},
reflecting a non-flat ES. For a fixed region $A$, $S_n$ has a
form of $\kc{1+1/n}a+b$, which appears to be in agreement with
the Cardy-Calabrese formula of Renyi entropy up to a constant.
While the constant $b$ depends on the length of $A$. 

Our strategy is applicable to other tensor constraints constructed
by tensor chains.

\begin{figure}
    \newcommand{\figheightTN}{73pt}
    \newcommand{\figheightTC}{28pt}
    \begin{minipage}[t]{0.31\linewidth}
        \centering
        \includegraphics[height=\figheightTN]{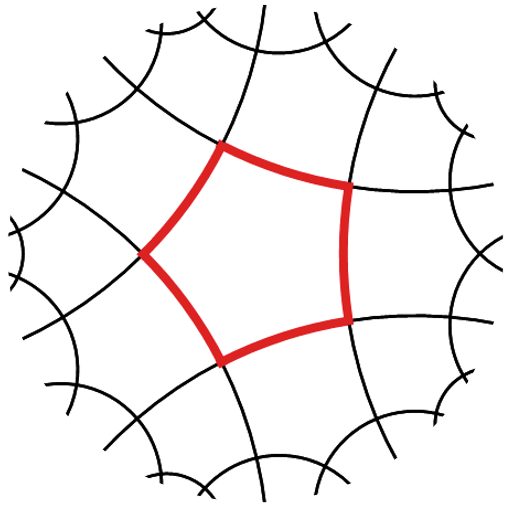}\\
        \includegraphics[height=28pt]{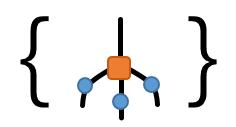}\\
        \caption{A tensor network with $\{5,4\}$ tiling and its tensor constraint. $\kappa_c=1/1=1$.
        }\label{Fig541}
    \end{minipage}
    \hfill
    \begin{minipage}[t]{0.65\linewidth}
        \centering
        \includegraphics[height=\figheightTN]{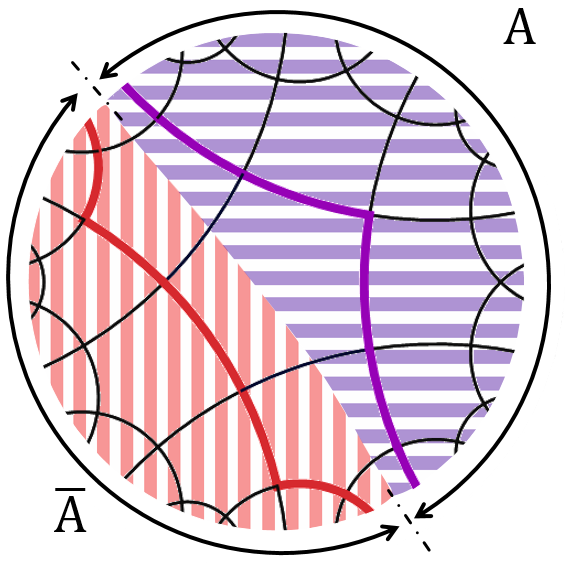}
        \includegraphics[height=\figheightTN]{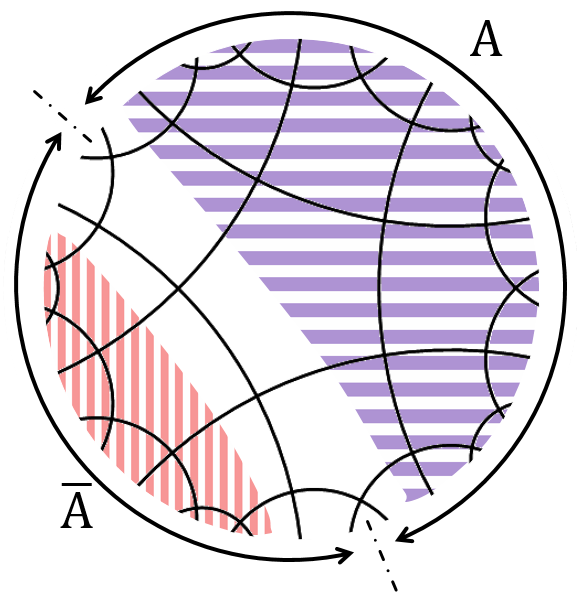} \\
        \includegraphics[height=\figheightTC]{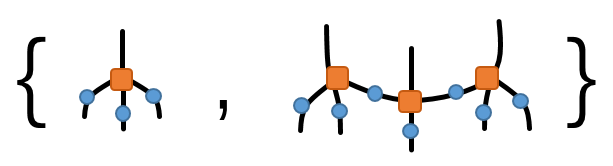}\\
        \caption{A tensor network with $\{5,4\}$ tiling and its tensor constraints. $\kappa_c=(1+2+2)/3=5/3$. Under the greedy algorithm, whether all the tensors are absorbed or not depends on the choice of $A$ and $\bar A$.
        }\label{Fig54111}
    \end{minipage}
\end{figure}

\begin{description}
        \item[$\kappa_c=1$.]  The tensor network with single constraint is plotted in Fig.~\ref{Fig541}.
Irrespective of the interval $A$ one picks out on the
boundary, no tensor is absorbed by the greedy algorithm. The
CP tensor chain is closed and we always obtain a non-flat ES. On
the other hand, wherever an operator is inserted in the bulk, it
can not be pushed to the boundary with the use of the isometry. So
such a tensor network does not enjoy QEC.

        \item[$\kappa_c=5/3$.] The tensor network with the constraint composed of three $T$ tensors is plotted in Fig.~\ref{Fig54111}. Given an interval $A$ on the boundary, an operator inserted in
        the wedge of $A$ can be pushed to $A$. So such a tensor
        network enjoys QEC. While, it is subtle to justify whether the
        ES is flat or not. We find both flat and
        non-flat ES can be obtained, which depends on the specific choice of the interval $A$, as shown in Fig.~\ref{Fig54111}. So we call this tensor network has a
        mixed ES.
    \end{description}
The constructions of tensors $T$ and $E$ in above two tensor networks are given in Appendix \ref{SectionExistence}.

We list the properties of entanglement for above tensor
networks with $\ke{5,4}$ tiling in ordering of their $\kappa_c$ in
Table.~\ref{TableCP}. We find that the higher $\kappa_c$ is, the
stronger is the ability of QEC, but the ES more easily becomes
flat. We remark that such a relation still holds in general
cases. A detailed analysis on tensor networks with general
tiling and general constraints is given in
\cite{Ling:2018ajv}, where the tensor networks with general
constraints in terms of tensor chains are classified based on
their properties of QEC and ES, with the power of CP reduced
interior angle $\kappa_c$. The four tensor networks considered
in this paper are typical examples of their own class.

\begin{table}
    \begin{tabular}{|c|c|c|c|c|}
        \hline
        $\kappa_c$ & 1 & 3/2 & 5/3 & 2 \\
        \hline
        QEC & N & Y & Y & Y \\
        \hline
        ES  & non-flat & non-flat & mixed & flat \\
        \hline
    \end{tabular}
    \caption{The entanglement properties of tensor networks. }\label{TableCP}
\end{table}

\section{Correlation function}

Taking the tensor network with constraint (\ref{5411PITC}) as an
example, we show that the two-point correlation function in
CFT can be reproduced here.

Given a local operator $O$ on the boundary, we may calculate the two-point correlation function
\begin{eqnarray}\label{Correlation}
    &&C(x_1,x_2)    \\
    &=&\frac{\bra{\Psi}O(x_1)O(x_2)\ket{\Psi}}{Z}
    -\frac{\bra{\Psi}O(x_1)\ket{\Psi}\bra{\Psi}O(x_2)\ket{\Psi}}{Z^2}. \nn
\end{eqnarray}

During the course of evaluating $\bra{\Psi}O(x_1)\ket{\Psi}$
($\bra{\Psi}O(x_2)\ket{\Psi}$), all of the indexes are contracted
except the indexes located at $x_1$ ($x_2$). It turns out
that except those tensors in the neighborhood of $x_1$ ($x_2$), most of other tensors are absorbed by the greedy algorithm.

\begin{figure}
    \centering
    \includegraphics[width=\linewidth]{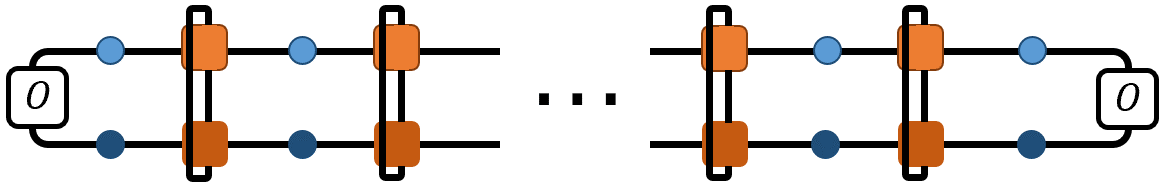}
    \caption{Those tensors which are not absorbed by the greedy algorithm in $\bra{\Psi}O(x_1)O(x_2)\ket{\Psi}$ form a bracket of MPS, where each $TT^\dagger$ pair denotes a site.}\label{FigMPS}
\end{figure}

In $\bra{\Psi}O(x_1)O(x_2)\ket{\Psi}$, all the indexes are
contracted except the indexes located at $x_1$ and $x_2$. The
greedy algorithm functions similarly as the case when we discuss
QEC and ES. Let us consider $x_1$ and $x_2$ as those marked
points in Fig.~\ref{Fig5411}, then the tensors which are not
absorbed by the greedy algorithm are just illustrated as in
Fig.~\ref{Fig5411}. As a result, the survived tensors $T$ and
$E$ in $\bra{\Psi}O(x_1)O(x_2)\ket{\Psi}$ form a bracket of matrix
product state (MPS) sandwiching $O(x_1)O(x_2)$, as shown in
Fig.\ref{FigMPS}.

Furthermore, the MPS is formed by the tensors along the
geodesic connecting $x_1$ and $x_2$. The length of the geodesic is
proportional to the number of tensor pairs $TT^\dagger$ in
Fig.\ref{FigMPS}, {\it i.e.} the number of sites of the MPS, which
is denoted as $l(x_1,x_2)$. Because of the tiling of $H^2$ space, when
two points are far from each other, we have \begin{eqnarray}\label{Distance}
    l(x_1,x_2) = c \log|x_1-x_2|,
\end{eqnarray}
where $|x_1-x_2|$ is the number of the indexes between $x_1$ and $x_2$ on the boundary and $c$ is a constant based on the tiling.

Based on the interpretation of MPS, when $|x_1-x_2|$ is large enough, we can expect that the correlation function behaves like
\begin{eqnarray}\label{CorrelationPowerLaw}
C(x_1,x_2) \sim e^{-ml(x_1,x_2)} = |x_1-x_2|^{-mc},
\end{eqnarray}
where the positive coefficient $m$ reflects the gap of the theory describing the MPS. Our interpretation from MPS shares the same strategy with the one from bulk field dynamics in \cite{Susskind:1998dq}. After all, (\ref{CorrelationPowerLaw}) agrees with the result in CFT.

With the specific construction of tensors $T$ and $E$ in
Appendix \ref{SectionExistence}, one can derive $C(x_1,x_2)$
concretely. In Appendix \ref{SectionCalculateCorrelation}, by
adopting the construction in Fig.~\ref{FigTE5411}, we show that
$C(x_1,x_2)$ satisfies (\ref{CorrelationPowerLaw}) indeed.
Especially, $m$ is determined by the inner construction of tensor
$T$ and $E$ as well as the type of the operator $O$.

\begin{figure}
    \newcommand{\minipagewidth}{0.48\linewidth}
    \newcommand{\figheight}{120pt}
    \begin{minipage}[t]{\minipagewidth}
        \centering
        \includegraphics[height=\figheight]{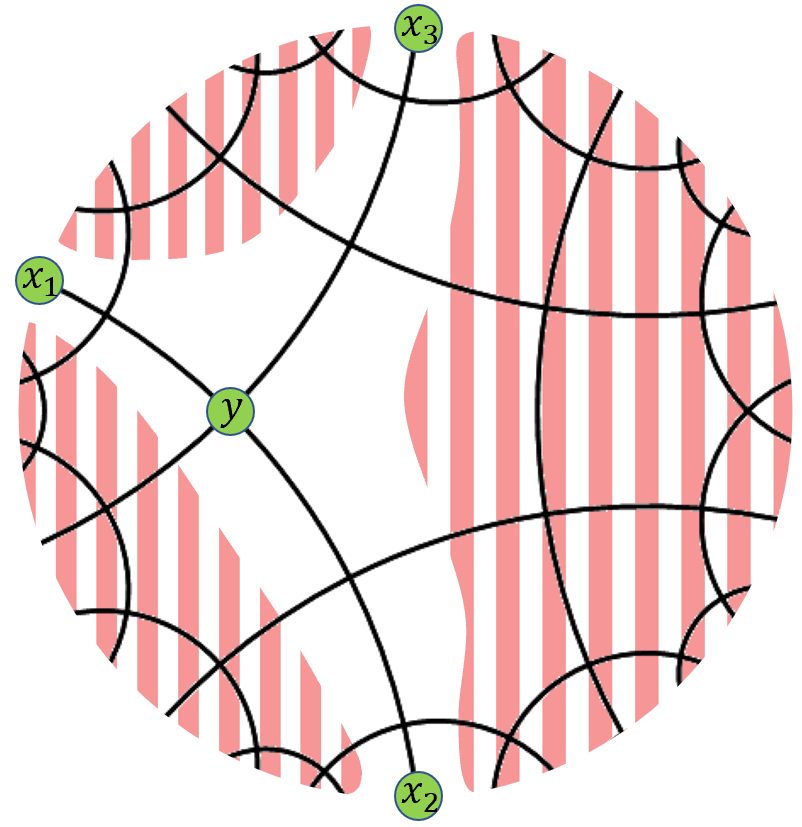}
        \caption{Three-point function.}\label{Fig3pt}
    \end{minipage}
    \hfill
    \begin{minipage}[t]{\minipagewidth}
        \centering
        \includegraphics[height=\figheight]{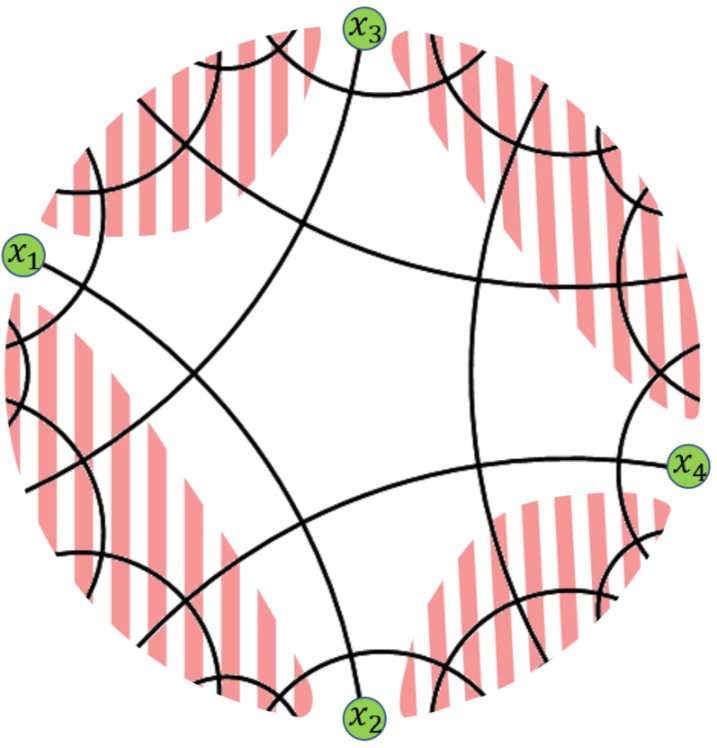}\\
        \caption{Four-point function.}\label{Fig4pt}
    \end{minipage}
\end{figure}

Similarly, those higher-point functions can be evaluated in
tensor networks as well. The network structure of three-point
function is simplified under the action of the greedy algorithm
into a MPS-like form: three linear MPSs are connected at a point
in the bulk $y$, as shown in Fig.~\ref{Fig3pt}. The three-point
correlation $C(x_1,x_2,x_3)$, characterized by the connected
part of $\bra\Psi O(x_1)O(x_2)O(x_3)\ket\Psi$, is supported by the
two-point correlations of MPS between $\ke{x_i,y}$ for
$i=1,2,3$ in the bulk. Thus,
\begin{eqnarray}\label{3pt}\begin{split}
    &C(x_1,x_2,x_3) \\
    \sim& \exp\ke{-m\kd{l(x_1,y)+l(x_2,y)+l(x_3,y)}}    \\
    \sim& \exp\ke{-\frac12 m\kd{l(x_1,x_2)+l(x_2,x_3)+l(x_3,x_1)}}          \\
    =&\kc{|x_1-x_2||x_2-x_3||x_3-x_1|}^{-mc/2},
\end{split}\end{eqnarray}
where (\ref{Distance}) is applied at the last step. (\ref{3pt}) agrees with the
result in CFT as well. Nevertheless, the network structure of
four-point function can not be simplified into a MPS-like form
any more, as shown in Fig.~\ref{Fig4pt}. A block of tensors in
the bulk prevents a geometrical estimation of correlation. It also agrees with the fact that conformal symmetry can not fully determine the form of four-point function in CFT.

Above analysis can be applied to other tensor networks. The greedy algorithm plays a similar role as in the evaluation of ES, except that the boundary effect will be suppressed by the
insertion of operators on the boundary.

Multi-point correlation functions in tensor networks can be
reduced into some brackets of MPS. For a general tensor network,
the generated MPS may have multiple layers. The number of
layers is approximately proportional to the distance between
two CP tensor chains beside the geodesic. It is interesting
to notice that given a tiling of tensor network, this distance
becomes larger with the increase of $\kappa_c$, which is observed if we compare different tensor networks in this paper with each other and is proved in \cite{Ling:2018ajv}. Such tendency implies that the correlation between two endpoints carried by the MPS becomes stronger as well.

\section{Conclusion and Outlook}
In this paper the notion of critical protection based on tensor
chain has been proposed to describe the behavior of tensor
networks under the action of greedy algorithm. In particular, a
criteria has been developed with the help of the average reduced
interior angle of CP chain such that for a given tensor network
the ability of QEC and the flatness of ES can be justified in a
quantitative manner. Currently it is still challenging to
construct tensor networks which could capture all the holographic
features of AdS spacetime. What we have found in this paper has
shed light on this issue. Firstly, we have learned that the notion
of critical protection provides a description on the limit of
information transmission with full fidelity. CP tensor chain
is the maximal boundary which can holographically store the
interior information \cite{Flammia:2016xvs,Jacobson:2015hqa}.
Thus, for a tensor network which is desired to capture the
feature of QEC as AdS space, it must not contain circular CP
curves. As a result, the tensor network with $\kappa_c=1$ in
this paper is not a candidate of holography. Furthermore, among the examples of tensor networks considered in this paper, the tensor network with $\kappa_c=3/2$ has more likelihood to mimic the AdS holography since it exhibits both features of QEC and non-flat ES, which motivates us to propose some strategy to construct tensor networks in more general setup which could capture the desirable holographic aspects of AdS space, which will be explored in \cite{Ling:2018ajv}.

The correlation in tensor networks with constraints becomes
more transparent since the greedy algorithm reduces the structure of network into MPS lying on the geodesic. The number of layers in MPS is determined by CP tensor chains. This fact can be understood as the realization of Witten diagram in AdS space.
Furthermore, from the viewpoint of field theory in bulk, the
correlation function is the partition function of a particle in
AdS space. Since the classical trajectory of the particle is just
the geodesic, the partition function has the same form as
(\ref{CorrelationPowerLaw}), where $m$ is the mass of the
particle. Therefore, we expect that the MPS may effectively
describe the trajectory of a particle in AdS space, where the
number of layers in MPS corresponds to the quantum fluctuations of the trajectory near the geodesic.

The geometric description of CP tensor chain is appealing. In the
light of its periodic structure, we find the analogy of CP tensor
chain is the curve of constant curvature in $H^2$ space such that
$\kappa_c$ is related to the geodesic curvature of the curve
\cite{Ling:2018ajv}. Specifically, an open CP tensor chain
corresponds to a hypercircle, which has a constant distance from
its axes (a geodesic), as illustrated in Fig.~\ref{Fig5411}. Such
a distance measures the deviation from RT formula when evaluating
the Renyi entropy, which may be linked to the tension of cosmic
brane in \cite{Dong:2016fnf}.

Because of the chain structure of tensor constraint, in our
present framework we have investigated QEC and ES only for a
single interval on the boundary. It is an open question whether
these properties of entanglement can be realized for
multi-intervals on the boundary, as investigated in network with
perfect tensors or random tensors
\cite{Hayden:2016cfa,Yang:2015uoa,Pastawski:2015qua}.

Finally, beyond the applications in holography, we expect that
tensor network models in this paper may be applicable to describe the quantum states of critical system in condensed matter physics as well, because of the $SL(2,R)$ symmetry in $H^2$ space.
Imposing tensor constraints in terms of tensor chains leads to a generalized greedy algorithm, which is completely under control and would greatly simplify the calculation involved in tensor networks. The correlation and entanglement of the tensor network state will be determined by the tiling style, the tensor constraints ($\kappa_c$) as well as the specific construction of elementary tensors.

 \medskip

    {\it We are grateful to Long Cheng, Glen Evenbly, Wencong Gan, Muxin Han, Ling-Yan Hung, Shao-Kai Jian, Hai Lin, Wei Li, Fuwen Shu, Yu Tian, Menghe Wu,
    Xiaoning Wu and Hongbao Zhang for helpful discussions and correspondence. This work
    is supported by the NSFC under
    Grant No. 11575195. Y.L. also acknowledges the support from
    Jiangxi young scientists (JingGang Star) program and 555 talent
    project of Jiangxi Province. Z.Y.X. is supported by National Postdoctoral Program for Innovative Talents BX20180318.}

\appendix

\section{Specific Construction of Tensors Subject to Tensor Constraints}\label{SectionExistence}

\begin{figure}
    \newcommand{\figheight}{0.15\textheight}
    \begin{minipage}[t]{0.47\linewidth}
        \centering
        \includegraphics[width=1.1\linewidth]{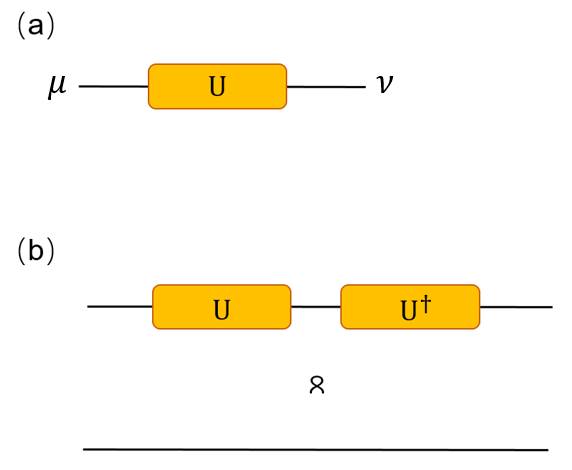}
        \caption{(a) Tensor $U$. (b) Tensor $U$ is proportional to an isometry.}\label{FigU}
    \end{minipage}
    \hfill
    \begin{minipage}[t]{0.47\linewidth}
        \centering
        \includegraphics[width=0.87\linewidth]{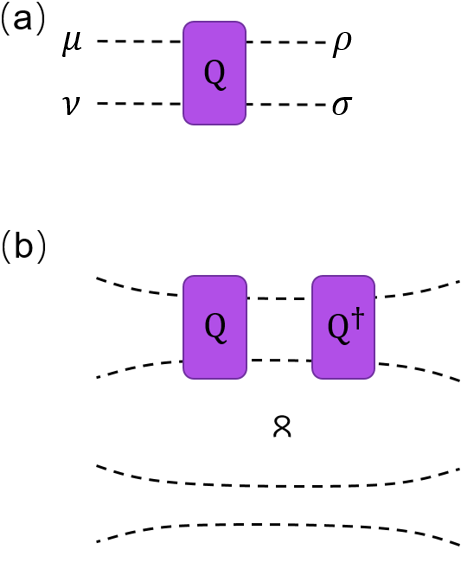}
        \caption{(a) Tensor $Q$, where two indexes on each side are grouped together. (b) Tensor $Q$ is proportional to an isometry between two grouped indexes.}\label{FigQ}
    \end{minipage}
    \hfill
    \begin{minipage}[t]{\linewidth}
        \centering
        \includegraphics[width=0.85\linewidth]{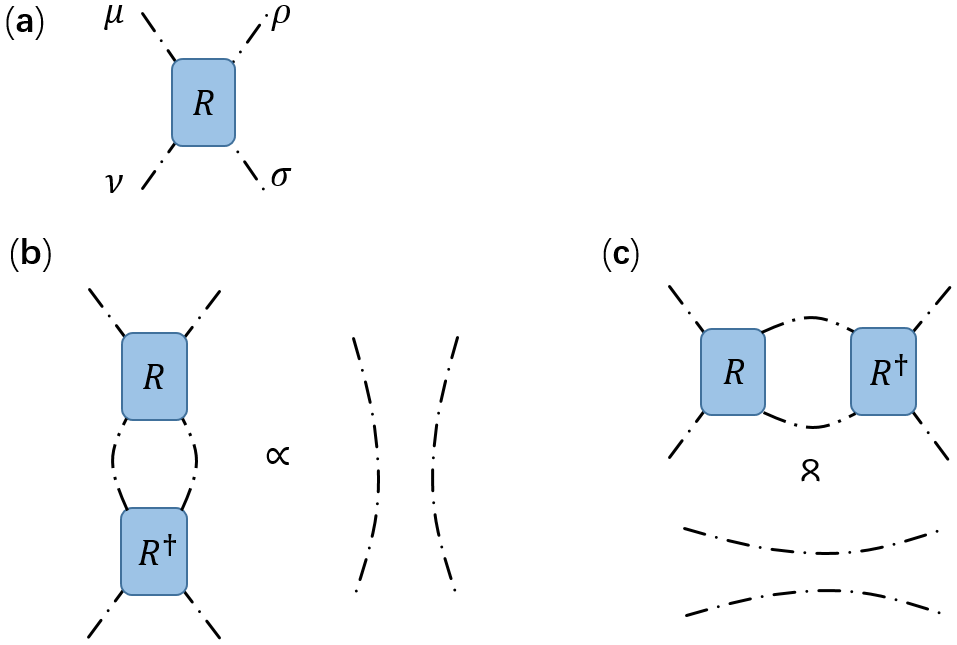}
        \caption{(a) Tensor $R$. (b) Tensor $R$ is proportional to isometries along two directions.}\label{FigR}
    \end{minipage}
\end{figure}

In Fig.\ref{FigU}, \ref{FigQ} and \ref{FigR},
we define tensor $U$, tensor $Q$ and tensor $R$ as the building blocks for $T$ and $E$. The elements of tensor $U$ are $U_{\mu\nu}$. They
satisfy following relations
\be\begin{split}
    U_{\mu\nu}=U_{\nu\mu}, \\
    \sum_{\nu} U_{\mu\nu}U_{\rho\nu}^*\propto\delta_{\mu\rho}.
\end{split}\ee
The elements of tensor $Q$ are $Q_{\mu\nu\rho\sigma}$ where
two indexes $\mu\nu$ ($\rho\sigma$) are grouped together.
They satisfy
\be\begin{split}
    Q_{\mu\nu\rho\sigma}=Q_{\rho\sigma\mu\nu}=Q_{\nu\mu\sigma\rho}, \\
    \sum_{\rho\sigma} Q_{\mu\nu\rho\sigma}Q_{\mu'\nu'\rho\sigma}^*\propto\delta_{\mu\mu'}\delta_{\nu\nu'}.
\end{split}\ee
The elements of tensor $R$ are $R_{\mu\nu\rho\sigma}$. They
satisfy
\be\begin{split}
    R_{\mu\nu\rho\sigma}=R_{\rho\sigma\mu\nu}=R_{\nu\mu\sigma\rho}, \\
    \sum_{\rho\sigma} R_{\mu\nu\rho\sigma}R_{\mu'\nu'\rho\sigma}^*\propto\delta_{\mu\mu'}\delta_{\nu\nu'}, \\
    \sum_{\nu\sigma} R_{\mu\nu\rho\sigma}R_{\mu'\nu\rho'\sigma}^*\propto\delta_{\mu\mu'}\delta_{\rho\rho'}.
\end{split}\ee

\begin{figure}
    \newcommand{\minipagewidth}{0.49\linewidth}
    \newcommand{\figheight}{0.09\textheight}
    \begin{minipage}[t]{\minipagewidth}
        \centering
        \subfigure[]{
            \includegraphics[height=\figheight]{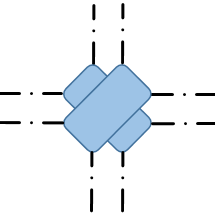}
        }
        \subfigure[]{
            \includegraphics[height=\figheight]{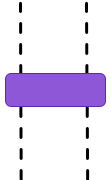}
        }
        \caption{(a) Tensor $T$ and \\ (b) tensor $E$ in (\ref{542PITC}) and Fig.\ref{Fig542}.}\label{FigTE542}
    \end{minipage}
    \hfill
    \begin{minipage}[t]{\minipagewidth}
        \centering
        \subfigure[]{
            \includegraphics[height=\figheight]{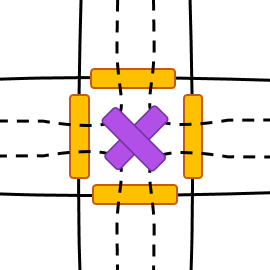}
        }
        \subfigure[]{
            \includegraphics[height=\figheight]{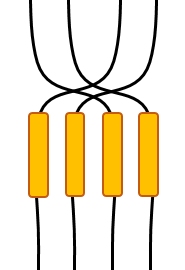}
        }
        \caption{(a) Tensor $T$ and \\ (b) tensor $E$ in (\ref{5411PITC}) and Fig.\ref{Fig5411}.}\label{FigTE5411}
    \end{minipage}
    \hfill
    \begin{minipage}[t]{\minipagewidth}
        \centering
        \subfigure[]{
            \includegraphics[height=\figheight]{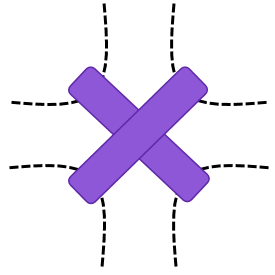}
        }
        \subfigure[]{
            \includegraphics[height=\figheight]{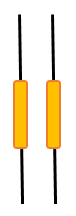}
        }
        \caption{(a) Tensor $T$ and \\ (b) tensor $E$ in Fig.\ref{Fig541}).}\label{FigTE541}
    \end{minipage}
    \hfill
    \begin{minipage}[t]{\minipagewidth}
        \centering
        \subfigure[]{
            \includegraphics[height=\figheight]{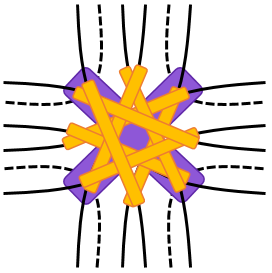}
        }
        \subfigure[]{
            \includegraphics[height=\figheight]{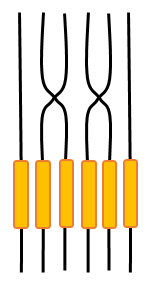}
        }
        \caption{(a) Tensor $T$ and \\ (b) tensor $E$ in Fig.\ref{Fig54111}).}\label{FigTE54111}
    \end{minipage}
\end{figure}

Specifically, we construct the tensor $T$ and tensor $E$ for the tensor network with $\ke{5,4}$ tiling for
different tensor constraints, as shown in
Fig.\ref{FigTE542},\ref{FigTE5411},\ref{FigTE541} and
\ref{FigTE54111}. Specific elements of some tensors $Q$ and $R$
are given in \cite{Evenbly:2017htn}.

\section{Correlation Function in a Specific Tensor Network}\label{SectionCalculateCorrelation}

\begin{figure*}
    \centering
    \includegraphics[width=0.8\linewidth]{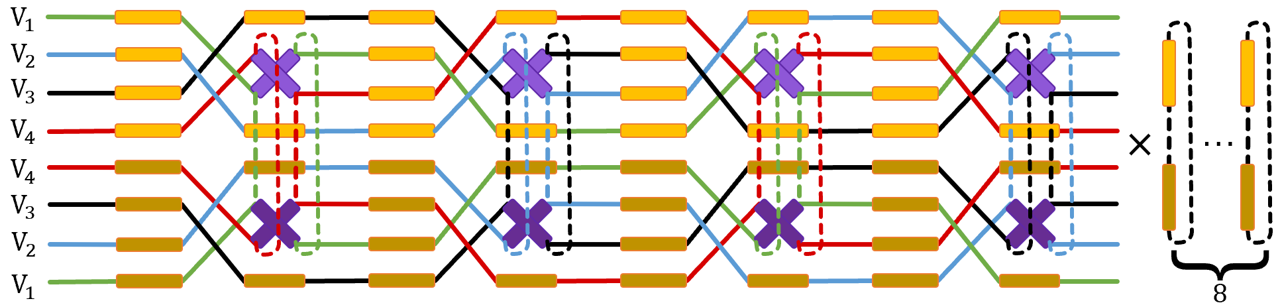}
    \caption{The inner structure of $V$, a part of $\tilde\rho$ containing $4$ sites. $V$ is an outer product of $4$ channels $\ke{V_1,V_2,V_3,V_4}$.}\label{FigInnerMPS}
\end{figure*}

Given a tensor network, we define the unnormalized reduced
density matrix of two points $\ke{x_1,x_2}$ on the boundary as
\begin{eqnarray}
\tilde\rho = \Tr_{\complement\ke{x_1,x_2}} \ket\Psi\bra\Psi,
\end{eqnarray}
where $\complement\ke{x_1,x_2}$ is the supplementary of two points
$\ke{x_1,x_2}$. Now we treat $\tilde\rho$ as a matrix with
column index in $\mathcal H_{x_1}\otimes \mathcal H_{x_1}^*$ and row index in $\mathcal H_{x_2}\otimes \mathcal H_{x_2}^*$, where superscript ``$*$'' refers to the dual space. Meanwhile, operator $O$ is treated as a column vector. Then all the brackets in (\ref{Correlation}) can be expressed in terms of matrix product, such as $\bra{\Psi}O(x_1)O'(x_2)\ket{\Psi}=O^T\tilde\rho
O'$. Thanks to greedy algorithm, $\tilde\rho$ has a form of MPS, as shown in Fig.~\ref{FigMPS}. One can show that $\tilde\rho$ is a symmetric matrix. Those tensors absorbed by the greedy algorithm contribute to $\tilde\rho$ as a constant factor but do not affect the correlation function, so we just set it to be $1$.

Adopting the specific construction of tensors in
Fig.~\ref{FigTE5411}, we demonstrate one part of the inner
structure of $\tilde\rho$ in Fig.~\ref{FigInnerMPS}, which plays a key role in the evaluation of the reduced density matrix. We
observe that such a network which looks complicated is actually an
outer product of $4$ individual networks and each of them has a
period composed of $4$ sites, where each site is denoted by
a pair of tensors $TT^\dagger$. We write $l(x_1,x_2)$ as $l$ for short. Since we are interested in the
behavior of correlation function at large scale, it is enough
to consider $l/4\in\mathbb Z$. Then $\tilde\rho$ can be decomposed into
\begin{eqnarray}\label{Decomposition}
    \tilde\rho=V^{l/4} F,\quad V=\bigotimes_{i=1}^{4} V_i, \quad F=E\otimes E^\dagger.
\end{eqnarray}
Four $V_i$'s have been marked out with different colors in
Fig.~\ref{FigInnerMPS}. Actually, $\ke{V_2,V_3,V_4}$ can be
obtained by cycling $V_1$. So they share the same eigenvalues.

To evaluate these eigenvalues explicitly, we set $U$ and $Q$
to be
\begin{eqnarray}\label{UQ}
U_{\mu\nu}=\delta_{\mu\nu},\quad Q_{\mu\nu\rho\sigma}=\sqrt{1-r^2}\delta_{\mu\rho}\delta_{\nu\sigma}+ir\delta_{\mu\sigma}\delta_{\nu\rho},
\end{eqnarray}
where $\mu,\nu,\rho,\sigma=1,2,\cdots,d$ and $0<r<1$. The case of $r=0,1$ should be excluded, since it leads to the flatness of ES.
Plugging it into $V_1$, we have
\begin{eqnarray}
(V_1)_{\mu\nu\rho\sigma}=d^3(1-r^4)\delta_{\mu\nu}\delta_{\rho\sigma}+d^4r^4\delta_{\mu\rho}\delta_{\nu\sigma},
\end{eqnarray}
which is real and symmetric. It can be diagonalized as
\begin{eqnarray}
V_1&\to& d^{4}\text{~diag}\ke{1,\underbrace{r^4,r^4,\cdots,r^4}_{d^2-1}},
\end{eqnarray}
where the eigenvector of the eigenvalue $d^4$ is $\delta_{\mu\nu}$. We can further diagonalize $V$ as
\begin{eqnarray}
V&\to&\bigotimes^4 V_1  \\
&\to& d^{16}\text{~diag}\left\{1,
    \underbrace{r^4,\cdots,r^4}_{4(d^2-1)},
    \underbrace{r^8,\cdots,r^8}_{6(d^2-1)^2},\right. \nn    \\ && \left.
    \underbrace{r^{12},\cdots,r^{12}}_{4(d^2-1)^3},
    \underbrace{r^{16},\cdots,r^{16}}_{(d^2-1)^4}\right\}.
\end{eqnarray}
The first eigenvector is the identity operator $I$. We use
$\alpha=0,1,2,3,4$ to label the degenerate subspace of eigenvalue $d^{16}r^{4\alpha}$. Since $\tilde\rho$ and $F$ are symmetric, from (\ref{Decomposition}), we have
\begin{eqnarray}
V F=F V^T.
\end{eqnarray}
So $F$ is diagonal between different subspaces.

We decompose the operator $O$ according to these five subspaces
\begin{eqnarray}
    O=\sum_{\alpha=0}^4 O_\alpha, \quad O_\alpha V=VO_\alpha=d^{16}r^{4\alpha} O_\alpha.
\end{eqnarray}
We find that
\begin{eqnarray}
I^TFI&=&d^4,    \\
O_\alpha^T FI=I^T FO_\alpha&=& g \delta_{\alpha 0}, \\
O_\alpha^T FO_{\alpha'}&=&f_\alpha\delta_{\alpha\alpha'},
\end{eqnarray}
where coefficients
\begin{eqnarray}
f_\alpha=O_\alpha^T FO_\alpha, \quad g^2=f_0 d^4.
\end{eqnarray}
Now those brackets in (\ref{Correlation}) can be evaluated.
\begin{eqnarray}
Z=& I^T\tilde\rho I=&d^{4l+4},\\
\bra{\Psi}O(x_1)\ket{\Psi}=&O^T\tilde\rho I=&d^{4l} g,  \\
\bra{\Psi}O(x_2)\ket{\Psi}=&I^T\tilde\rho O=&d^{4l} g,  \\
\bra{\Psi}O(x_1)O(x_2)\ket{\Psi}=&O^T\tilde\rho O
=&\sum_{\alpha=0}^4 d^{4l}r^{\alpha l}f_\alpha.
\end{eqnarray}
Finally,
\begin{eqnarray}
C(x_1,x_2)
&=&\frac{\sum_\alpha d^{4l}r^{\alpha l}f_\alpha}{d^{4l+4}}
-\kc{\frac{d^{4l} g}{d^{4l+4}}}^2   \\
&=&\sum_{\alpha=1}^4 f_\alpha d^{-4}r^{\alpha l}    \\
&=&\sum_{\alpha=1}^4 f_\alpha d^{-4}e^{-(\alpha\log\frac1{r}) l}
\end{eqnarray}
which behaves like (\ref{CorrelationPowerLaw}) with $m=\alpha\log\frac1{r}>0$ for the minimal $\alpha$ such that $f_\alpha\neq0$.

\end{document}